# First Directions for Using Gamification to Motivate for Open Access

Athanasios Mazarakis[1] and Paula Bräuer[1]

[1] Kiel University, Kiel 24118, Germany
`a.mazarakis@zbw.eu, p.braeuer@zbw.eu`

**Abstract.** Most scientists are aware that, in addition to the traditional and subscription-based publication model, there is also the possibility of publishing their research in open access. Various surveys show that scientists are in favour of this new model. Nevertheless, the transition to open access has been very slow so far. In order to accelerate this process, we are looking for new opportunities to create incentives for researchers to deal with the topic of open access. In a field study with 28 participants the effects of the game design elements badge and progress bar on the motivation when working on an online quiz on the topic of open access are examined. In our study both game design elements provide a statistically significant increase in the number of questions answered compared to a control group. This suggests that gamification is useful to motivate for open access.

## 1 Introduction

Many universities, research institutions and funding organisations are now advocating open access, i.e. open access for scientific literature, but the adaption so far has been rather slow [29]. This situation is even more crucial as funding agencies like the European Union demand open access for scientific publications and research data, which are funded under the Horizon 2020 programme. Article 29.2 of the Model Grant Agreement, which is binding for all participants of the Horizon 2020 program, states this explicitly [9].

Scientists in particular have a decisive influence on the transformation of publication models to open access through their choice and decision where to publish their research. Many scientists and scholars have a fundamentally positive view of the idea of free availability of their research, in particular for publicly funded research results. Nevertheless, this approach is yet not very widespread [18]. In order to reward researchers for making their results openly available and to motivate them to do so, incentives must be applied [6]. We are convinced that gamification could be an appropriate motivation method to engage scientists to get information about open access. Based on this assumption is the expectation to create incentives for them to publish open access in the long-term.

Gamification refers to the use of game design elements in a context other than that of a game [8]. The concept of gamification has meanwhile established itself in the field of human-computer interaction [31]. In various application cases from fitness and health [20, 28] to working environments [19, 32, 36] to education and training [4, 17] it could be shown that gamification has a positive influence on increasing motivation and performance. But how gamification can be used to motivate scientists, who mostly work under special conditions and have special criteria for the evaluation of their performance, has not yet been sufficiently researched [11].

Usually work can have the potential to motivate on its own, especially if a flow condition is achieved [7]. Gamification is a method to create or promote a flow condition [15]. But in contrast to flow theory, which requires full concentration to get into a flow condition [7], gamification works with both conscious and unconscious perceived game design elements.

Additionally it is necessary to explore open access because scientists are interested in their research and reputation but not so much in different aspects of publishing. Even worse, researchers want to focus on other things and not on open access, because they consider such a topic like publishing issues as complex and boring [21]. Yet, as already shown, this topic is of highest importance for all scientists who publish their publicly funded results [9].

The aim of this article is to examine the impact of the game design elements badge and progress bar on researchers in the context of an open access quiz. This will make a first contribution to filling the research gap about gamification and open access. The results of our preliminary study, which is part of a larger study, provide information on whether gamification is suitable for encouraging researchers to engage with the topic of open access.

The rest of our article is structured as follows: An overview of the current state of research is given in the following section. Section three describes the study in detail, giving information about the procedure of our experiment, measurements and the subjects which participated. The results follow in the fourth section. In the last section, the results are discussed and pointed to limitations and possible future research.

## 2  Related Work

Gamification is used in science to promote participation in citizen science projects such as Galaxy Zoo or Foldit [3, 10, 30]. Through the use of game design elements, incentives are created so that users generate new solutions or classify images and not only once but preferably on a long-term basis.

Kidwell and colleagues provide empirical evidence how badges can motivate researchers to publish scientific results according to open science principles. The authors showed that the award of open science badges motivates researchers to make their research data freely available [22].

Scientific platforms such as ImpactStory and ResearchGate also use game design elements such as badges or progress indicators to motivate scientists. Hammarfelt and colleagues [16] look at the impact of gamification on these platforms. The authors suspect that the use of game design elements creates a motivating feeling when publications and online interactions with other users are transformed into points. They also assume that their own position within the academic community can be more easily defined through gamification. Scientific progress and comparison with other users is thus facilitated by converting publications into points [16].

Feger and colleagues address the question of how gamification can be used in science workplaces [11]. They present problems and challenges, such as the aspect that progress in scientific work is difficult to quantify. In a follow-up study, based on an extensive survey of high energy physicists, they developed various gamified approaches to motivate researchers to provide reproducible data [12].

For the present study, the two game design elements badge and progress bar were selected in order to examine their effects individually and thus assess their potential to encourage scientists to deal with open access. Our study is a preliminary study of a larger study with the aim of improving the impact of gamification as a tool for disseminating and improving knowledge about open access. Both game design elements are used to build nonfiction gamified experiences [37] and are easier to reuse in other contexts than narratives, for example. This is in line with our approach, as we are looking for possibilities of gamification that can also be used in subsequent studies on open access.

Badges in the context of games and gamification are digital artefacts which are given to the user for the fulfilment of certain tasks and are therefore a visual representation [1]. The game design element badge has already proven itself in other studies as an effective instrument for increasing motivation [14, 27]. It can therefore be assumed that badges can also be used effectively in the context of open access to motivate academics for this topic.

In contrast to badges, the progress bar element is a less frequently studied game design element in gamification [23, 34]. It is a simple visual way of informing the users about their progress. Initial studies have shown that even this simple form of feedback can motivate users [27]. Besides games, this easily implemented element is also used in many other areas. These include, for example, software to show a progress in an installation process or in surveys to give the participant an overview of how many questions he or she still has to answer. Despite the general assumption that progress bars should increase the completion rate of surveys, there are also contrary results which should not be ignored [24].

## 3    Method

The aim of our study is to investigate whether scientists can be motivated by gamification to deal with the topic of open access. To this end, a field study in the form of an online quiz on the topic of open access was developed. A between-subjects design was chosen to determine the effect of the individual game design elements. The following three experimental conditions were used for the experiment: a control group (CG) without gamification, an experimental group with progress bar (PB) and an experimental group with badges (BA).

### 3.1 Hypotheses

In line with previous findings [14, 26, 27] the following two hypotheses were formulated:

**H1**: The subjects in the group with progress bar (PB) answer on average more open access questions than the control group (CG).

**H2**: The subjects in the group with badges (BA) answer on average more open access questions than the control group (CG).

### 3.2 Procedure

To answer our hypotheses we developed an experimental setting with a quiz. The quiz is based on the idea that questions are only answered as long as the participants enjoy it. This has been explicitly stated in the invitation e-mail and at the first page of the experiment. There is no minimum number of questions to answer. Instead the quiz can be finished at any time. Participants are randomly and permanently assigned to one of the three test conditions (CG, BA or PB). All subjects reached the online quiz through the same internet URL. Due to the random division into groups, the subjects were not aware that there is more than one test condition. At the end of the experiment, the mean number of questions answered is compared between the three groups.

The quiz developed for the field study consists in total 29 multiple-choice questions on the topic of open access. Each question is provided with four possible answers of which only one is correct. In all three test conditions, the subjects are immediately informed of the correct or incorrect answer to a question. The correct answer is highlighted in green and wrong answers are displayed in red (see Figure 1). All questions are always asked in the same order, so that we can compare the results between the different groups.

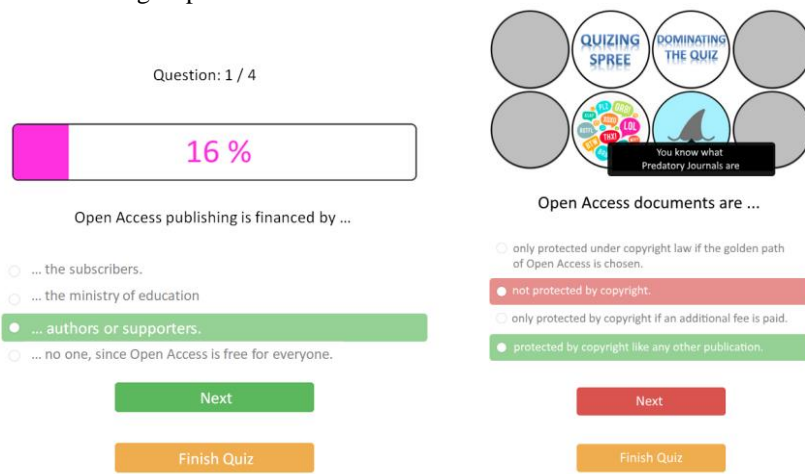

**Figure 1.** Correctly answered question in the progress bar condition (left) and incorrectly answered question in the badge condition (right).

All questions deal with different aspects of open access. In a pre-test, five subjects were asked to give feedback about our online quiz. All questions were described as "interesting" and "important". At this point, the questions were tested exclusively with the non-gamified version of the quiz. This form of presentation of the questions was found to be "monotonous" by the subjects. This supports the findings of Kelty [21] and provides the prerequisites for using gamification to increase motivation to engage users for a significantly longer period of time with the topic.

The badges are displayed to the user above the questions during the quiz (see Figure 1 right). Before a badge is unlocked, only a grey circle is visible in the overview. So the participants can estimate that further badges can be achieved. This can appeal to the collective instinct [37] and thus motivate to continue with an activity. The badges were designed in such a way that they can be unlocked as consistently as possible in the course of the experiment. The subjects are not informed in advance about the criteria for the award of the badges. This was done on the purpose not to encourage the subjects to just unlock the badges instead of dealing extensively with the questions. In addition, unexpected unlocking can have a motivating effect on people who react positively to surprise effects [37]. Just after a badge is unlocked the condition for its assignment is disclosed and the associated visual badge is displayed.

Eight different badges were designed for the quiz. One of the badges is bound to a time condition and is activated for answering a question within five seconds. Three of the badges are awarded for a certain number of questions answered correctly one after the other. Three more badges are given for correctly answering certain individual questions and the last badge is assigned for answering all questions in the quiz.

A simple horizontal progress bar was used for the other game design element (see Figure 1 left). However, the progress bar is not filled equally, but in uneven proportions. This procedure is based on the assumption that the maximum number of questions cannot be predicted easily and that this vague approach might motivate the subjects [25]. If the maximum number of questions is known, the subjects might be additionally motivated to answer more questions, which would lead to a confusion of the results, as this would set an implicit goal. This would make it impossible to attribute our results only to the game design elements because the ceteris paribus assumption would not be applicable due to the intervening factor of the additional variable, more specifically the implicit goal. It is also known from research on the influence of progress indicators on the completion rate of surveys that they can also be demotivating if they give the participant the feeling he or she is not progressing fast enough [5]. The variation in progress should add a random factor that is more motivating than a monotonous increase in the display and still reflect progress.

At the beginning of the quiz the subjects were also asked to indicate their age and gender and after completing the quiz they were given the opportunity to leave a comment in a text field. The subjects did not receive any compensation for their participation in this experiment.

## 4  Results

Over a period of one week, 28 participants were acquired at three German institutes who are also project partners of the forthcoming larger study. Only employees of the institutes were contacted and no students. They received an e-mail with the request to participate in a study and could access the quiz anonymously by clicking on a general link to the page where the quiz was provided. Both in the e-mail and on the website the subjects were informed that they only had to answer questions for as long as they enjoyed it. A total of three volunteers did not report their gender. Of the other subjects, 12 were male and 13 female. The mean age was 38.57 years (standard deviation 13.90). The experiment presented here is a preliminary study of a larger study. This explains the short runtime and the relatively small number of subjects.

On average, 23.32 of 29 maximum possible questions were answered (standard deviation 9.24). 18 subjects (64 %) answered all 29 questions, of which three were in the control group (corresponding to 27 % in this group), six in the group with badges (100 %) and nine in the group with progress bar (81 %). The evaluation thus shows a ceiling effect that is particularly pronounced in the two experimental conditions. The discussion can already be anticipated and additional questions should be developed for further use of the open access quiz. This can also be seen from the text field comments of the subjects, which at the end of the experiment could be submitted on a voluntary basis. Several respondents from all three groups noted that they would have liked to have answered further questions.

A list of the number of subjects per test condition, as well as the number of those who answered all questions, the mean value of answered questions and the associated standard deviation are given in Table 1.

**Table 1.** Total number of respondents, number of respondents with all answered questions, mean value and standard deviation per condition.

|  | Condition | | |
|---|---|---|---|
|  | CG | PB | BA |
| N | 11 | 11 | 6 |
| All answered | 3 | 9 | 6 |
| Mean | 16.82 | 26.73* | 29.00** |
| SD | 10.00 | 7.21 | 0.00 |

*= $p < .05$; ** = $p < .01$

An analysis of variance is performed for statistical evaluation. The test to check the homogeneity of the variances (Levene test) for the number of answered questions yields a statistically significant result with $p = .001$, the Levene statistics is 8.78. All following results are therefore based on an unequal variance and the corrected results are reported accordingly.

The analysis of variance yields a statistically significant difference between the individual test conditions, $F(2, 25) = 6.47$, $p = .005$. Since the homogeneity of the variances is not given, the Welch test must correct accordingly. The comparison of the mean values of the control group with the group with progress bar provides a statisti-cally significant result, $t(18.19) = 2.67$, $p = .008$. Consequently, the hypothesis H1 can be supported and it can be assumed that the progress bar motivated the subjects to answer more questions.

The comparison of the mean values between the control group and the experimental group with badges also shows a statistically significant result, $t(10.00) = 4.04$, $p = .001$. It can thus also be supported hypothesis H2 that the game design element badge also motivated the subjects to answer further questions.

The number of subjects in the individual test conditions differs significantly and, in addition, the number of subjects in the test conditions is very low. This can result in conservative effects in their statistical significance due to a low test power [13]. Therefore, it is all the more remarkable that statistically significant results have been obtained. This is also shown by the effect size of $\Delta = .99$ for the comparison between the control group and the progress bar. The effect size for the comparison between the control group and the group with badges is $\Delta = 1.22$. Thus a strong effect can be demonstrated for both comparisons. For such a small sample size and the problems normally associated with this size [13], the results are particularly remarkable.

## 5      Discussion and Future Work

The present study was able to show that the game design elements progress bar and badge had a motivating effect in the context of an open access quiz. Based on the results of Kidwell et al. [22], the assumption can thus be supported that the game design elements can also be used in a different context in order to generate or increase motivation for the topic of open access.

Despite the successful implementation of the field study, there are some limitations. First of all the small number of subjects is basically a problem for every field study. In the present context this can be explained because the study was scheduled to take one week to complete and that it is a preliminary study of a larger study.

The restriction that the data were collected exclusively in Germany could also have had an influence on the results. The level of knowledge and implementation of open access certainly makes a distinction in some cases between Germany and other countries. Also cultural factors may have an influence on the impact of gamification [35].

In addition, for further investigations of different game design elements with the same research design, more questions need to be developed to avoid the ceiling effect present in this study, which occurred especially in the experimental conditions. This is also in line with the comments from our subjects.

Another point that should be further examined is whether a cooperative design of the game design elements would be preferable in the scientific context. According to Feger et al. [12], the design of gamified scientific work environments should rather be based on a cooperative approach, as the scientists surveyed seemed to prefer such a design. Although the present study did not include a cooperative element we achieved very positive results. In order to be able to demonstrate possible advantages of cooperative design methods, an experimental investigation of cooperative gamification approaches would have to be carried out in the scientific context.

Also it should be examined whether an online quiz with additional questions is too artificial to approach the question of how open access can be made more motivating. Instead, another way could be to let subjects make fictitious selection decisions in an experiment (conjoint analysis), similar to a procedure described by Schöbel and colleagues [33]. In such an experiment, for example, scientific articles could be sorted in an order in which they would be prioritized for download. Here, different aspects of open access could be visualised and highlighted by gamification. A further possibility for experimental research could be to offer different publication options for one's own publication and to evaluate these according to the probability of submission by the subjects. The variants could differ with regard to the number, design and quality of game design elements and bibliometric data.

Finally, another aspect that will be considered in the following studies is the impact of possible differences in game design elements, especially in the context of open access, on researchers from different disciplines and career levels. On the one hand, there are huge differences in publication behaviour between the various scientific disciplines [2, 29]. On the other hand, it can be assumed that the level of knowledge and attitudes towards open access varies greatly. This in turn could have an influence on how gamification works in this context and how it should be designed.

Further studies will examine these aspects. The present study has already shown that it is possible to engage subjects about open access by means of a gamified quiz. Based on the available results, we will conduct a follow-up study to investigate the long-term effects of the two applied game design elements in an online community of practice of researchers. This will be investigated including further demographic data, information on the current position and the currently active research discipline.

**ACKNOWLEDGMENTS**

This research was funded by the Federal Ministry of Education and Research (BMBF) in Germany as part of the research project OA-FWM (16OA044C).